\documentclass[a4paper,11pt]{article}
\usepackage{color}
\usepackage{graphicx}
\usepackage{amssymb}
\usepackage{bm}
\usepackage{amsmath,color}
\newcommand{\T}{\mathcal{T}}
\newcommand{\E}{\varepsilon}
\newcommand{\de}{\partial}
\newcommand{\ket}[1]{\left| #1 \right>} 
\newcommand{\bra}[1]{\left< #1 \right|} 
\newcommand{\vev}[1]{\langle #1 \rangle}

\begin{document}

\setcounter{page}{0} \topmargin0pt \oddsidemargin0mm \renewcommand{%
\thefootnote}{\fnsymbol{footnote}} \newpage \setcounter{page}{0}
\begin{titlepage}
\vspace{0.2cm}
\begin{center}
{\Large {\bf Composite branch-point twist fields in the Ising model and their expectation values}}

\vspace{0.8cm} {\large \text{Emanuele Levi$^{\star}$}}

\vspace{0.2cm}
{ Centre for Mathematical Science, City University London, \\
Northampton Square, London EC1V 0HB, UK}\\
\end{center}
\vspace{1cm}

We investigate a particular two-point function of the $n$-copy Ising model. That is, the correlation function $\vev{\E(r)\T(0)}$ involving the energy field and the branch-point twist field.
The latter is associated to the symmetry of the theory under cyclic permutations of its copies. We use a form factor expansion to obtain an exact integral representation of $\vev{\E(r)\T(0)}$
and find its complete short distance expansion. This allows us to identify all the fields contributing in the short distance massive OPE of the correlation function under examination, and fix their
expectation values, conformal structure constants and massive corrections thereof. Most contributions are given by the composite field $:\E\T:$ and its derivatives. We find all
non-vanishing form factors of this latter operator.


 \vfill{
\hspace*{-9mm}
\begin{tabular}{l}
\rule{6 cm}{0.05 mm}\\
$^\star \text{emanuele.levi.1@city.ac.uk}$\\
\end{tabular}}

\end{titlepage}
\newpage

\section{Introduction}
The idea of associating a local field to a branch point in a CFT on a Riemann surface was first introduced in \cite{riemann},
and deepened in \cite{cardycalabrese}. It became then natural in the context of the evaluation of bi-partite entanglement entropy for
two-dimensional integrable QFTs (see i.e. \cite{entropy} for a review), and for non-integrable ones \cite{ben}. These models are usually
quantized on a line, and then evolved in euclidean time $\tau=it$. The idea is to divide the system at $\tau=0$ into two regions, and then
consider $n$ copies of the original model. These ``replicas'' are connected cyclically along a branch cut which covers one of the two regions
(the choice is indeed arbitrary). The boundary points of the cut are branch points of degree $n$, and a field $\T$ is associated to them,
related to the $\mathbb{Z}_{n}$ symmetry of the global model. These fields are called twist fields, and are local with respect to the
hamiltonian density of the $n$-copy model $h_{n}(x)=\sum_{j=1}^{n}h_{j}(x)$, where $h_{j}(x)$ is evaluated in the $j$-th sheet of the
$n$-root map which connects the Riemann surface with $\mathbb{R}^{2}$. Consider for simplicity  $\mathcal{M}_{0}$ the manifold with a
single branch point in $x=0$ and $\mathbb{R}^{2}_{n}$ the disconnected $n$ replicas, then the action of the twist field on the expectation
values of the theory is defined as
\begin{equation}
\vev{\ldots}_{\mathcal{M}_{0}}=\frac{\vev{\T(0)\ldots}_{\mathbb{R}^{2}_{n}}}{\vev{\T}_{\mathbb{R}^{2}_{n}}}.
\label{firstdef}
\end{equation}
The counterpart of the twist field in the ultraviolet conformal theory is a spinless operator of conformal dimension \cite{cardycalabrese}
\begin{equation}
    \Delta_{\mathcal{T}}=\bar{\Delta}_{\mathcal{T}}=\frac{c}{24}\left(n-\frac{1}{n}\right),
    \label{known1}
\end{equation}
In a recent paper \cite{us1} involving the author, it was first observed that a function that generalizes $\Delta_\T$ to massive theories, known in literature as $\Delta$-function \cite{DSC} has all the right properties to be a
$c$-function of the theory. That is it has the same qualitative features of Zamolodchikov's $c$-function \cite{Zamc}, even if it is different from it. This observation was put on a more solid basis in \cite{us2}, where the correlation function $\vev{\phi(t)\T(0)}$ between the perturbing field and the
twist field in general $1+1$-dimensional integrable theories was considered. In particular it was noticed that it can be obtained exactly for the $n$-copy thermally perturbed
Ising model, where $\phi=\E$, that is the energy field of conformal dimension $\Delta=\bar{\Delta}=1/2$. This gives us the opportunity of investigating the massive OPE for the operator
product $\E(r)\T(0)$ in great detail. This is the main issue we want to address in this paper.\\
The present work is organized as follows: after Section \ref{intr}, that is devoted to describe the model and set the notation, the analysis of the two point function $\vev{\E(r)\T(0)}$ and the identification of the operator $:\E\T:$ is carried out in Section \ref{OPE}. The two-particle form factor of this composite operator is found in Section \ref{composite}, and higher-particle form factors in Section \ref{higher}. In Section \ref{conclusions} we collect all our main conclusions, and discuss open problems.

\section{The Ising model and twist field form factors}
\label{intr}
The two dimensional thermally perturbed Ising model can be described by a relativistic quantum field theory on a line with action
\begin{equation}
S_E=i\int d^{2}x \left(  \bar{\psi}(\de_x-\de_t) \bar{\psi}-\psi(\de_x+\de_t) \psi-m\bar{\psi}\psi  \right),
 \label{action}
\end{equation}
which can be obtained by perturbing the critical Ising model with the energy operator and introducing a mass scale $m$ which is related to the temperature.
While in the conformal theory this field can be taken to be the product of two fermionic fields, in the off critical model, due to the presence of $m$, there is an additional mixing with the identity operator, so that 
\begin{equation}
\label{ambig}
\E=a \bar{\psi}\psi+ bm \mathbb{I}\hspace{2cm} \text{with}\hspace{.5cm} a,b\in \mathbb{R}\setminus \{0\}.
\end{equation}
In (\ref{action}) $\psi$ and $\bar{\psi}$ are the two real components of a Majorana spinor, and they can be represented as
\begin{equation}
\begin{split}
 \psi(\tau,x)&=\sqrt{\frac{m}{4\pi}}\int d\theta e^\frac{\theta}{2}\{ a(\theta)e^{m(ix\sinh \theta-\tau\cosh \theta)}+a^{\dagger}(\theta) e^{-m(ix\sinh \theta-\tau\cosh \theta)}\}\\
 \bar{\psi}(\tau,x)&=-i\sqrt{\frac{m}{4\pi}}\int d\theta e^{-\frac{\theta}{2}}\{ a(\theta)e^{m(ix\sinh \theta-\tau\cosh \theta)}-
a^{\dagger}(\theta) e^{-m(ix\sinh \theta-\tau\cosh \theta)}\},
\label{psipsibar}
\end{split}
\end{equation}
where $\theta$ is the rapidity, and the mode operators $a(\theta)$ and $a^\dagger(\theta)$ satisfy the canonical commutation relations
\begin{equation}
  \{ a(\theta) , a^{\dagger}(\theta') \} = \delta (\theta-\theta')\hspace{1cm}\{ a(\theta) , a(\theta') \} = \{ a^{\dagger}(\theta) , a^{\dagger}(\theta') \}=0.
\label{antia}
\end{equation}
The Hilbert space of this model is the Fock space built over the postulated vacuum $\ket{0}$ and the algebra (\ref{antia}). The action of the
operators $a$ and $a^\dagger$ is
\begin{equation}
a(\theta)\ket{0}=\bra{0} a^\dagger (\theta)\equiv 0 \hspace{1cm} a^\dagger (\theta)\ket{0}\equiv\ket{\theta} \hspace{1cm}
\bra{0}a(\theta)\equiv\bra{\theta},
\label{creann}
\end{equation}
and a general vector is defined by multiple applications of (\ref{creann})
\begin{equation}
\ket{\theta_1,\theta_2,...,\theta_n}=a^\dagger(\theta_1)a^\dagger(\theta_2)...a^\dagger(\theta_n)\ket{0},
\label{genstate}
\end{equation}
where a basis is chosen setting $\theta_1>\theta_2>...>\theta_n$.\\

The fields of the theory act as linear operators on the Hilbert space. For an operator $\mathcal{O}$ the general matrix element
 \begin{equation}
F_k^{\mathcal{O}}(\theta_1,\dots,\theta_k)=\bra{0}\mathcal{O}(0)\ket{\theta_1,\dots,\theta_k}_{in},
  \label{formfact1}
 \end{equation}
is known in literature as a form factor\footnote{in general form factors depend on a complete set of quantum numbers. The one-copy Ising model though has only one kind of ``particle'' in its
spectrum, so that a form factor needs no other specifications than rapidities of incoming particles, whereas for the $n$-copy theory there is an additional number that has to be specified that labels the copy the ``particles'' belong to.}. These quantities are particularly useful in the context of integrable theories,
and they provide a non-perturbative expansion for two-point functions. Indeed by the introduction of the resolution of the identity over the basis (\ref{genstate}) it is possible to
express the correlator between two operators $\mathcal{O}_{1}$ and $\mathcal{O}_{2}$ as\footnote{throughout this paper the following notation will be adopted $\mathcal{O}(\tau=0,x)\equiv\mathcal{O}(x)$}
\begin{equation}
\label{nonpert}
\vev{\mathcal{O}_{1}(r)\mathcal{O}_{2}(0)}=\sum_{k=1}^{\infty
}\int\limits_{\theta_1>\theta_2>\cdots>\theta_k }\frac{d\theta _{1}\ldots d\theta _{k}}{(2\pi)^{k}}
F_{k}^{\mathcal{O}_{1}}(\theta_1,\dots,\theta_k)\left(F_k^{\mathcal{O}_{2}}(\theta_1,\dots,\theta_k)\right)^{*}e^{-mr\sum_{i=1}^{k}\cosh\theta_{i}}.
\end{equation}
The convergence of this series is remarkably fast for most integrable models, and leads to a good approximation just considering the first few terms.
The aforementioned constants $a$ and $b$ in (\ref{ambig}) are fixed by the normalization  
\begin{equation}
\label{normaldefinition}
 \E(x)=2\pi\left(m \mathbb{I} + :\bar{\psi}(x)\psi(x): \right),
\end{equation}
where $aa^\dagger=:aa^\dagger:+\mathbb{I}$ is the definition of normal ordering in this case. 
The two-particle form factor can be easily extracted using (\ref{psipsibar})
\begin{equation}
\begin{split}
F_2^\E(\theta_1,\theta_2)&=\bra{0}\E(0)\ket{\theta_1\theta_2}=-i\frac{m}{2}\int d\phi d\eta e^{\frac{\phi-\eta}{2}}
\bra{0}a(\eta)a(\phi)a^\dagger(\theta_1)a^\dagger(\theta_2)\ket{0}=\\
&=- i m \sinh \frac{\theta_1-\theta_2}{2},
\end{split}
\label{e2ff}
\end{equation}
where the Wick theorem for fermion algebrae is used to define contractions. The normalization (\ref{normaldefinition}) is chosen in light of the relation between $\E$ and the trace of the stress energy tensor $\Theta(x)=2\pi m \E(x)$ \cite{Zamolodchikov:1989zs}, to give $F_2^\Theta(i\pi)=2\pi m^2$.

\subsection{Form factors construction for $\T$}

The twist field is a field in the general meaning of the term: is an object for which expectation values and exchange relations with all
 the other fields in the theory can be defined. In particular its action on the fermion fields is
\begin{eqnarray}
    \psi_{i}(\tau,y)\mathcal{T}(\tau,x) &=& \mathcal{T}(\tau,x) \psi_{i+1}(\tau,y) \qquad x^{1}> y^{1}, \nonumber\\
    \psi_{i}(\tau,y)\mathcal{T}(\tau,x) &=& \mathcal{T}(\tau,x) \psi_{i}(\tau,y) \qquad x^{1}< y^{1}, \label{exchange}
\end{eqnarray}
where $\psi_{i}$ is the fermionic field of the $i^{\text{th}}$ copy, and the same relations hold for $\bar{\psi}_{i}$.\\
These non-trivial relations modify Watson's equations
\begin{eqnarray}
  F_{k}^{\mathcal{T}|n_1\dots n_i  n_{i+1} \dots }(\dots,\theta_i, \theta_{i+1}, \dots ) &=&
  S_{n_i n_{i+1}}^{(n)}(\theta_{i\,i+1})
  F_{k}^{\mathcal{T}|\dots n_{i+1}  n_{i} \dots}(\dots,\theta_{i+1}, \theta_i,  \dots ),
  \label{1}\\
 F_{k}^{\mathcal{T}|n_1 n_2 \dots n_k}(\theta_1+2 \pi i, \dots,
\theta_k) &=&
  F_{k}^{\mathcal{T}| n_2 \dots n_k n_1+1}(\theta_2, \dots, \theta_{k},
  \theta_1),\label{2}
  \end{eqnarray}
where particles' indexes $n_i$  label the copy they belong to, $\theta_{i\,i+1}=\theta_i-\theta_{i+1}$ and, for the Ising model,
 $S_{n_i n_{i+1}}^{(n)}(\theta_{i\,i+1})=(-1)^{\delta_{n_i,n_{i+1}}}$.\\
The pole structure is also modified by (\ref{exchange}) which introduces an extra kinematic equation as follows
  \begin{eqnarray}
 \begin{array}{l}
\\
  \text{Res}  \\
 {\footnotesize \bar{\theta}_{0}={\theta}_{0}}
\end{array}\!\!\!\!
 F_{k+2}^{\mathcal{T}|n_i n_i  n_1 \dots n_k}(\bar{\theta}_0+i\pi,{\theta}_{0}, \theta_1, \dots, \theta_k)
  &=&
  i \,F_{k}^{\mathcal{T}| n_1 \dots n_k}(\theta_1, \ldots,\theta_k). \label{3}
  \\
\begin{array}{l}
\\
  \text{Res}  \\
 {\footnotesize \bar{\theta}_{0}={\theta}_{0}}
\end{array}\!\!\!\!
 F_{k+2}^{\mathcal{T}|n_i (n_i+1) n_1 \dots n_k}(\bar{\theta}_0+i\pi,{\theta}_{0}, \theta_1, \dots, \theta_k)
  &=&-i\prod_{j=1}^{k} S_{n_in_j}^{(n)}(\theta_{0j})
  F_{k}^{\mathcal{T}| n_1 \dots n_k}(\theta_1, \dots,\theta_k),\label{kre}
\end{eqnarray}
These equations connect form factors of different orders, and can be used to check the validity of any form factors associated to the twist field.\\
The two particle form factor for the Ising model was found in \cite{entropy2}, and can be expressed as
\begin{equation}
F_{2}^{\T|11}(\theta)=\frac{i\vev{\T}\cos\left(\frac{\pi}{2n}\right)}{n}\frac{\sinh\left(\frac{\theta}{2n}\right)}{\sinh
\left(\frac{\theta+i\pi}{2n}\right)\sinh\left(\frac{\theta-i\pi}{2n}\right)},
\label{2pffising2}
\end{equation}
and the exact expression of $\vev{\T}$ was also given in that paper.
Moreover higher particle form factors were also extracted, and they showed a Pfaffian structure. As $\T$ is an even operator under the
$\mathbb{Z}_{2}$ symmetry of the Ising model only even particle form factors are non-vanishing, and they have the form
\begin{equation}\label{f}
F_{2k}^{\mathcal{T}|11\ldots 1}(\theta_1,
\ldots,\theta_{2k})=\langle\mathcal{T}\rangle{\text{Pf}}(K),
\end{equation}
where ${\text{Pf}}(K)^2 = \det(K)$ is a Pfaffian, and $K$ is an anti-symmetric $2k \times 2k$ matrix, with entries
\begin{equation}\label{k}
  K_{ij}=\frac{F^{\T|11}(\theta_{ij})}{\vev{\T}}.
\end{equation}

\section{An important OPE and composite twist fields}
\label{OPE}
 In \cite{us2} the authors were interested in the computation of the delta-function of the twist field
  \begin{equation}\label{delta}
     \Delta(r)=-\frac{n}{2}\int_{r}^{\infty} ds\, s\left( \frac{\langle\Theta(s)\T(0)\rangle_{\mathbb{R}^2_{n}}}{\vev{\T}_{\mathbb{R}^2_{n}}}-
\langle\Theta\rangle_{\mathbb{R}^2}\right),
 \end{equation}
 where $\Theta$ is the trace of the energy-momentum tensor.
There, evidence was given that this function has all the right properties to be a c-function. 
In light of the relation between $\Theta$ and $\E$ mentioned in Section \ref{intr} we will henceforth refer directly to this second operator.
Indeed for $\E$ only the two-particle form factor is non-vanishing and it is possible to find an exact
integral representation for the correlation function\footnote{from now on the subscript labelling the manifold VEVs are evaluated on will be omitted. It is implicit that $\vev{\T}$
is always evaluated on $n$ disconnected copies of $\mathbb{R}^2$, while other VEVs are on the real plane.}
\begin{equation}
\langle\E(r)\T(0)\rangle=\langle\E\rangle\vev{\T}-\frac{\vev{\T} m}{ 2\pi^{2}} \cos\frac{\pi}{2n} \int\limits_{-\infty }^{\infty } dx
\frac{K_0(2mr\cosh\frac{x}{2})\sinh\frac{x}{2n}\sinh\frac{x}{2}}{\cosh\frac{x}{n}-\cos\frac{\pi}{n}}
\label{integral1},
\end{equation}
where $K_0(\phi)$ is the modified Bessel function of the second kind of argument $\phi$.
In the CFT, where the correlation length tends to infinity, one is allowed to use the OPE\footnote{this OPE is as usual to be taken in the ``weak'' sense; indeed it has a meaning only
once plugged into a matrix element of the CFT on $\mathbb{R}^{2}_{n}$.}
 \begin{equation}
 \label{ope1}
    \E(r)\T(0)=\sum_{k}\tilde{C}_{\E \mathcal{T}}^{k}
 r^{2(\Delta_{k}-\frac{1}{2}-\Delta_\mathcal{T})}
  \mathcal{O}_{k}(0),
 \end{equation}
where $\mathcal{O}_{k}$ is a basis of fields, and $\tilde{C}_{\E \mathcal{T}}^{k}$ are the dimensionless constants of the expansion.
The most relevant operator appearing in (\ref{ope1}) is the composite twist field $\mathcal{O}_0\equiv:\E\T:$, and it is defined implicitly as the twist operator which corresponds to the leading
term. Its conformal weight and structure constant with $\E$ and $\T$ are known \cite{us2}
 \begin{equation}\label{weight}
   \Delta_{:\E \mathcal{T}:}=\frac{1}{2n}+\Delta_{\mathcal{T}}\quad\text{and}\quad
   \tilde{C}_{\E \mathcal{T}}^{:\E \mathcal{T}:}=\frac{1}{n}.
  \end{equation}
We know that, due to the arbitrariness in the choice of the argument of
\footnote{indeed the choice to take as argument 0 is arbitrary, and one could have chosen any point in the interval $[0,r]$. The difference between these choices is represented by a
Taylor expansion about $0$, therefore the OPE (\ref{ope1}) should include all derivatives of the field $:\E\T:$.}
$:\E\T:$, also derivatives of this field play a role in the OPE (\ref{ope1}).
We are able to fix the weight and structure constants for these corrections, and to identify 
$\mathcal{O}_\alpha\equiv:\de^{2\alpha}\E \T:$ with $\alpha\in \mathbb{N}$
\begin{equation}\label{weight2}
   \Delta_{:\de^{2\alpha}\E \mathcal{T}:}=\frac{1+2\alpha}{2n}+\Delta_{\mathcal{T}}\quad\text{and}\quad
   \tilde{C}_{\E \mathcal{T}}^{:\de^{2\alpha}\E \mathcal{T}:}=\frac{1}{n(2\alpha)!},
\end{equation}
where $\de^{2\alpha}\equiv(\de^{2}_{z\bar{z}})^{\alpha}$. Notice that since $\E$ and $\T$ are both spinless operators only these particular derivatives can contribute to the expansion. From now on
we will refer directly to the operators $:\de^{2\alpha}\E \mathcal{T}:$, denoting $:\E \mathcal{T}:$ as the case $\alpha=0$.\\

In the massive theory an analogue but different OPE exists \cite{Z} which can be extracted peturbatively from (\ref{ope1}), and can be used to express the LHS of (\ref{integral1}).
Here and throughout this paper the assumption that there be the same basis of operators in the massless and massive theories is pushed forward, such that we will refer with the same
symbols (with an abuse of notation) to the CFT and perturbed fields.
This expansion is in a sense richer than the conformal one, and contains information about the scaling region sorrounding the conformal fixed point. This reflects on the OPE by
giving a nonzero VEV to some of $\mathcal{O}_{k}$, which are of nonlocal nature. In fact while in the CFT these values are identically zero due to scale invariance,
 in the massive case they are $\vev{\mathcal{O}_{k}}\sim m^{2\Delta_{k}}$. The coefficients of this expansion contain information only about the short distance features of the theory, hence they are analytical with respect to the mass. Since in the Ising model the coupling constant is proportional to $m$, it is sensible to 
define an expansion of the form \cite{Z}
\begin{equation}
C_{\E \mathcal{T}}^{k}(r)=\tilde{\mathcal{C}}^{k}r^{2(\Delta_{k}-\frac{1}{2}-\Delta_\mathcal{T})}\left[1+\mathcal{C}_{1}^{k}mr\right.
+\left.\mathcal{C}_{2}^{k}(mr)^{2}+\dots\right].
\end{equation}
By means of (\ref{weight}) and (\ref{weight2}) we can express this expansion for the composite twist fields as
\begin{equation}
\label{structure}
C_{\E \mathcal{T}}^{:\de^{2\alpha}\E \mathcal{T}:}(r)=\frac{r^{\frac{1+2\alpha}{n}-1}}{n(2\alpha)!}\left[1+\mathcal{C}_{1}^{:\de^{2\alpha}\E \mathcal{T}:}mr\right.
+\left.\mathcal{C}_{2}^{:\de^{2\alpha}\E \mathcal{T}:}(mr)^{2}+\dots\right] \hspace{1.5cm}\text{for}\hspace{.5cm}\alpha \in \mathbb{N}_{0}.
\end{equation}
Fixing the perturbation constants appearing in (\ref{structure}) is generally a very hard task already for the first order. However, due to the special nature of the OPE under consideration we are able to determine all $\mathcal{C}_j^{:\de^{2\alpha}\E\T:}$ in a systematic way.
Therefore we expect the OPE in the massive theory to take the form
\begin{equation}
 \label{massiveope}
\E(r)\T(0)=\sum_{\alpha=0}^{\infty}\mathcal{C}_{\E\T}^{:\de^{2\alpha}\E\T:}(r):\de^{2\alpha}\E\T:(0),
\end{equation}
although we see later that further corrections to (\ref{massiveope}), that are not predictable from CFT arguments, will also appear.

\subsection{Computation of $\vev{:\de^{2\alpha}\E\T:}$ and $\mathcal{C}_{j}^{:\de^{2\alpha}\E\T:}$}
Let us now proceed by expanding 
the integral in the RHS of (\ref{integral1}) in a small $r$ region, and then compare the result with the massive OPE (\ref{massiveope}), that is we compare terms
with the same dimensions. In order to extract all the information needed it is convenient first to reexpress the integral in terms of the quantity $t=mre^{\frac{x}{2}}$
 \begin{equation}
 -\frac{ m}{ \pi^{2}} \cos\frac{\pi}{2n} (mr)^{\frac{1}{n}-1}
 \int\limits_{mr}^{\infty } dt\hspace{.3cm} t^{-\frac{1}{n}}\hspace{.3cm}K_0\left(t+\frac{(mr)^2}{t}\right)
 \frac{ \left[ 1 - \left( \frac{t}{mr} \right)^{-2} \right] \left[ 1 - \left( \frac{t}{mr} \right)^{ -\frac{2}{n} } \right] }{ \left( \frac{t}{mr} \right)^{ -\frac{4}{n} } - 2\cos\frac{\pi}{n} \left( \frac{t}{mr} \right)^{-\frac{2}{n}}+1}.
 \label{int2}
 \end{equation}
One can notice that leading contributions to this integral are given for large $t/(mr)$. This provides a natural parameter over which is possible to expand the fraction in (\ref{int2}) as
 \begin{equation}
  \frac{ \left[ 1 - \left( \frac{t}{mr} \right)^{-2} \right] \left[ 1 - \left( \frac{t}{mr} \right)^{ -\frac{2}{n} } \right] }{ \left( \frac{t}{mr} \right)^{ -\frac{4}{n} } - 2\cos\frac{\pi}{n} \left( \frac{t}{mr} \right)^{-\frac{2}{n}}+1}
  =\sum_{\alpha=0}^{\infty} \Omega_{\alpha}(n)\left(\frac{t}{mr} \right)^{-\frac{2\alpha}{n}},
 \label{exp1}
 \end{equation}
 where the coefficients $\Omega_{\alpha}(n)$ are real numbers, evaluated in Appendix \ref{A}.
Once (\ref{exp1}) is plugged into (\ref{int2}) one gets
 \begin{equation}
 -\frac{ m}{ \pi^{2}} \cos\frac{\pi}{2n}\sum_{\alpha=0}^{\infty} \Omega_{\alpha}(n)(mr)^{\frac{1+2\alpha}{n}-1}
 \int\limits_{mr}^{\infty } dt\hspace{.5cm} t^{-\frac{1+2\alpha}{n}}K_0\left(t+\frac{(mr)^2}{t}\right).
 \label{int3}
 \end{equation}
The dependence on $mr$ of the Bessel function can be extracted by expanding it for small $r$ as
 \begin{equation}
 K_0\left(t+\frac{(mr)^2}{t}\right)=K_0(t)-\frac{K_1(t)}{t}(mr)^{2}+\frac{K_{0}(t)+K_{2}(t)}{4t^{2}}(mr)^{4}+O(mr)^{6}.
 \label{besselexp}
 \end{equation}
The resulting integrals can be thought of as a special case of a known integral of the Maijer G-function, as explained in Appendix \ref{B}. Substituting (\ref{besselexp}) into (\ref{int3}) and carrying out the integrals gives 
\begin{equation}
\label{general}
\begin{split}
\langle&\E(r)\T(0)\rangle=-\frac{ \vev{\T}m}{ \pi^{2}} \cos\frac{\pi}{2n}\sum_{\alpha=0}^{\infty} \Omega_{\alpha}(n)\left[\frac{\Gamma \left( \frac{n-1-2\alpha}{2n} \right)^2}{2^{1+\frac{1+2\alpha}{n}}}(mr)^{\frac{1+2\alpha}{n}-1}\right.+ \\
&+\frac{n}{n+1+2\alpha}\frac{\Gamma \left( \frac{n-1-2\alpha}{2n} \right)^2}{2^{1+\frac{1+2\alpha}{n}}}(mr)^{\frac{1+2\alpha}{n}+1}+\left(\frac{n}{n+1+2\alpha}\right)^{2}\frac{\Gamma \left( \frac{n-1-2\alpha}{2n} \right)^2}{2^{2+\frac{2+2\alpha}{n}}}(mr)^{\frac{1+2\alpha}{n}+3}+\\
&-\left.  \frac{n^{3}}{(n+1+2\alpha)^{2}(3n+1+2\alpha)}\frac{\Gamma \left( \frac{n-1-2\alpha}{2n} \right)^2}{2^{2+\frac{2+2\alpha}{n}}}(mr)^{\frac{1+2\alpha}{n}+3}+\dots  \right],
\end{split}
\end{equation}
where we reported only the first contributions related to the composite twist field and its descendants, as we are mainly interested in those.
After a bit of manipulation we can compare term by term this expansion with (\ref{massiveope}), and by matching the terms with the same perturbative order we are able to extract the
following VEVs
\begin{equation}
 \label{vevs}
\vev{:\de^{2\alpha}\E\T:}=-\frac{\cos\frac{(1+2\alpha)\pi}{2n}(2\alpha)!n}{2^{1+\frac{1+2\alpha}{n}}\pi^2}
\Gamma \left( \frac{n-1-2\alpha}{2n} \right)^2 m^{\frac{1+2\alpha}{n}} \vev{\T}.
\end{equation}
This is the main result of this section.
In the same way we can fix the constants $\mathcal{C}_j^{:\de^{2\alpha}\E\T:}$ to every order.
A major challenge when doing this is to be able, for terms proportional to the same power of $r$, to distinguish between contributions to expectation values and to structure constants. It turns out that this ambiguity can be resolved by requiring that the expectation values (\ref{vevs}) are continuous functions of $n$ for each fixed value of $\alpha$. This requirement is natural because of the special relation between $\T$ and the entanglement entropy, in which context it is necessary to analytically continue all physical quantities to $n \in [1,\infty)$.
The first few non-vanishing coefficients are
\begin{equation}
 \label{coefficients}
\begin{split}
\mathcal{C}_2^{:\de^{2\alpha}\E\T:}=&\frac{n}{n+1+2\alpha}+ \frac{n^2}{(n+1+2\alpha)^2}\frac{\tan\frac{(1+2\alpha)\pi}{2n}}{\tan\frac{\pi}{2n}} \\
\mathcal{C}_4^{:\de^{2\alpha}\E\T:}=&  \frac{ n^2 }{2(n+1+2\alpha)(3n+1+2\alpha)}+\frac{n^3 (1+2 \alpha+2 n)}{(1+2 \alpha+n)^2 (1+2 \alpha+3 n)^2}
\frac{\tan\frac{(1+2\alpha)\pi}{2n}}{\tan\frac{\pi}{2n}}\\
\mathcal{C}_6^{:\de^{2\alpha}\E\T:}=&\frac{n^3(4n+1+2\alpha)}{6(n+1+2\alpha)^2(3n+1+2\alpha)(5n+1+2\alpha)}\\
&+\frac{\sin\frac{(1+2\alpha)\pi}{2n}}{\sin\frac{\pi}{2n}} \frac{n^4(4n+1+2\alpha)}{2(n+1+2\alpha)^2(3n+1+2\alpha)^2(5n+1+2\alpha)}
\left(1-2 n-\frac{2 n^2}{1+3 n+2 \alpha } \right).
\end{split}
\end{equation}
We find that $\mathcal{C}_{2j+1}^{:\de^{2\alpha}\E\T:}=0$, for $j\in \mathbb{N}_0$.\\
It is worth noticing that VEVs of these composite operators are singular whenever the argument of the Gamma-function is either zero or a negative integer. Analyzing (\ref{vevs}) one can see
that when $n=1+2\alpha$ the VEV of the $2\alpha^{\text{th}}$ derivative is divergent. In other words, such singularities can only occur for odd values of $n$.
Therefore, it follows that the analysis above is only really consistent when restricting $n$ to be even, in which case no singularities for special values of $n$ and $\alpha$ arise. 
In the $n$ odd case, the singularities that occur at various orders of the expansion actually cancel each other, generating a well defined short distance expansion. This stark contrast 
between the $n$ even and $n$ odd cases is a priori rather surprising. At present, we do not have a clear understanding of why it happens. It is a matter which we might came back to in 
future research. The leading term in (\ref{massiveope}) is well defined for all values of $n>1$, for both $n$ even and odd so that our identification of the expectation value of 
$:\varepsilon\mathcal{T}:$, already performed in \cite{us2}, does still hold for general values of $n$.  

\subsection{Logarithmic corrections to the massive OPE}
Analyzing carefully the contributions reported in Appendix \ref{B}, one can notice that further corrections involving terms of the form $(mr)^{2\alpha}$ and $(mr)^{2\alpha}\log(mr)$,  with 
$\alpha=0,1,\ldots$ occur (note for example the logarithmic terms in (\ref{finalexpansion1}) and (\ref{finalexpansion2})), such that (\ref{massiveope}) is modified to
\begin{equation}
\label{massiveope2}
\E(r)\T(0)=\sum_{\alpha=0}^{\infty}\left[\mathcal{C}_{\E\T}^{:\de^{2\alpha}\E\T:}(r):\de^{2\alpha}\E\T:(0)+m\mathcal{C}_{\E\T}^{\de^{2\alpha}\T}(r)\de^{2\alpha}\T(0)\right],
\end{equation}
where the new terms mentioned above are contained in the coefficients $\mathcal{C}_{\E\T}^{\de^{2\alpha}\T}(r)$. 
It is clear that they are features of the massive theory as they have no counterpart in the CFT. To understand their presence one has to consider once more (\ref{ambig}). 
The term proportional to $\bar{\psi}\psi$ is responsible for all the composite fields $:\de^{2\alpha}\varepsilon\mathcal{T}:$ in the OPE, whereas the term proportional to the 
identity generates contributions proportional to $m \mathcal{T}$ and its derivatives $m\de^{2\alpha}\T$. It is interesting that this paves the way for the evaluation
of $\vev{\de^{2\alpha}\T}$, although in this case, due to the structure of the expansion, this would need a resummation of infinitely many terms.
The presence of logarithmic terms is imputable to the freedom in choosing $b$ in (\ref{ambig}). Indeed in the expansions
carried out in Appendix B we have general terms of the type
\begin{equation}
 \label{logcorr}
(mr)^{2\alpha}\sum_{\beta=0}^{\infty}\left( \Upsilon(\alpha,\beta,n)+\Lambda(\alpha,\beta,n) \log mr \right),
\end{equation}
where $\Upsilon$ and $\Lambda$ are two rational functions of $\alpha$, $\beta$ and $n$. When a correction of the kind of (\ref{logcorr}) is plugged into (\ref{general}) it gives for fixed $\alpha$
\begin{equation}
\label{logcorr2}
-\frac{ \vev{\T}m}{ \pi^{2}} \cos\frac{\pi}{2n}(mr)^{2\alpha}\sum_{\beta=0}^{\infty}  \Omega_{\beta}(n)\left( \Upsilon(\alpha,\beta,n)+\Lambda(\alpha,\beta,n) \log mr \right).
\end{equation}
The term containing $\Upsilon$ contributes to $\vev{\E}\vev{\de^{2\alpha}\T}$. The presence of the logarithmic term allows us to rewrite (\ref{logcorr2}) as   
\begin{equation}
\label{logcorr3}
-\frac{ \vev{\T}m}{ \pi^{2}} \cos\frac{\pi}{2n}(mr)^{2\alpha}\sum_{\beta=0}^{\infty}  \Omega_{\beta}(n)\left( \Upsilon(\alpha,\beta,n)+\Lambda(\alpha,\beta,n) \log \delta+
\Lambda(\alpha,\beta,n) \log \frac{mr}{\delta} \right),
\end{equation}
where $\delta\in \mathbb{R}^+\setminus \{0\}$. This corresponds to a redefinition of $\vev{\E}$, and one can notice that a logarithmic correction is the only functional form that allows this to happen. This is not the first time a logarithmic correction to two point functions of the Ising model has been observed (see for example the spin-spin correlation function in \cite{spinspin}). In both cases their presence is fully explained by the ambiguity in the definition of $\vev{\E}$.

%
%


\section{Two particle form factors of composite operators}
\label{composite}
This section is dedicated to the evaluation of form factors for composite twist fields of the type introduced in Section \ref{OPE}.
These operators are formally defined as the regularised limit of an operator $\mathcal{O}$ approaching the twist field in the original CFT.
The regularisation defines the meaning of the ordered product, which can be taken to be a point splitting
\begin{equation}
\label{definition}
:\mathcal{O}\T:(x)\sim \lim_{\E \rightarrow 0} \mathcal{O}(x+\E)\T(x).
\end{equation}
One can start by considering $:\psi \T:$ as a benchmark. Since $\T$ is even and $\psi$ is odd under the $\mathbb{Z}_{2}$ symmetry of the Ising model, only odd particle form factors will be non-vanishing. Considering then the matrix element
$\bra{0}\psi(x)\T(0)\ket{\theta}$ the one-particle form factor can be extracted by looking at the leading contribution as $x$ approaches $0$. It is useful to introduce the resolution of the identity for the $n$-copies system
\begin{equation}
   {\mathbb{I}}=\sum_{k=1}^{\infty
}\sum_{\mu _{1}\ldots \mu _{k}}\int\limits_{\theta_1>\theta_2>\cdots>\theta_k }\frac{d\theta _{1}\ldots d\theta _{k}}{%
(2\pi
)^{k}}|\theta_1,\ldots,\theta_k\rangle_{\mu_1,\ldots,\mu_k}\,
   {\,}_{\mu_k,\ldots,\mu_1}\!\langle
   \theta_k,\ldots,\theta_1|\label{csum}
\end{equation}
where $\mu_{i}$ label the number of the copy and all quantum numbers of the $i$-th particle. In this way one can write
\begin{equation}
\bra{0}\psi(x)\T(0)\ket{\theta}=\frac{n}{2\pi}\int_{-\infty}^{\infty} d\phi \bra{0}\psi(x)\ket{\phi} \bra{\phi}\T(0)\ket{\theta}.
\label{0OTtheta}
\end{equation}
The matrix element $\bra{0}\psi(x)\ket{\phi}$ can be easily extracted from (\ref{psipsibar})
\begin{equation}
\bra{0}\psi(x)\ket{\phi}=\sqrt{\frac{m}{4\pi}} e^{\frac{\phi}{2}-ixp_{\phi}},
\label{0psiphi}
\end{equation}
while $\bra{\phi}\T(0)\ket{\theta}$ is linked to the two-particle form factor by crossing \cite{smirn}
\begin{equation}
\bra{\phi}\T(0)\ket{\theta}=\bra{0}\T(0)\ket{\theta,\phi+i\pi-i\E^{+}}+\vev{\T}\delta(\theta-\phi),
\label{crossing}
\end{equation}
where the introduction of $i\E^{+}$ has to be thought of in the distributions sense, and describes how to avoid the pole at $i\pi$.\\
Plugging (\ref{0psiphi}) and (\ref{crossing}) into (\ref{0OTtheta}) yields
\begin{equation}
\begin{split}
\bra{0}\psi(x)\T(0)\ket{\theta}=&\vev{\T}\sqrt{\frac{m}{4\pi}} e^{\frac{\theta}{2}-ixp_{\theta}}+\frac{i}{2\pi}\sqrt{\frac{m}{4\pi}}\vev{\T}\cos\frac{\pi}{2n}\\
&\int d\phi\hspace{.5cm} e^{\frac{\phi}{2}-ixp_{\phi}}\frac{\sinh\left(\frac{\theta-\phi-i\pi}{2n}\right)}{\sinh\left(\frac{\theta-\phi+i\E^{+}}{2n}\right)\sinh\left(\frac{\theta-\phi-2i\pi}{2n}\right)}.
\end{split}
\label{0OTtheta2}
\end{equation}
The leading term when $x$ approaches $0$ is given by the integral part,
that is divergent, but can be made convergent by shifting the domain of integration from the real axe $\mathbb{R}$ to $\mathbb{R}+i\pi/2$. Such a change of variable does not affect the
result of the integration, as the integrand has no poles in the region between the two axes\footnote{notice that it is the term $i\E^{+}$ which allows to perform this shift.}.\\
The integral in (\ref{0OTtheta2}) becomes then
\begin{equation}
e^{-i\frac{\pi}{4}} \int d\phi \hspace{.5cm}e^{\frac{\phi}{2}-xE_{\phi}}\frac{\sinh\left(\frac{\theta-\phi-i\pi/2}{2n}\right)}{\sinh\left(\frac{\theta-\phi+i\pi/2}{2n}\right)\sinh\left(\frac{\theta-\phi-i3\pi/2}{2n}\right)}.
\label{shift1}
\end{equation}
In this integral the main contributions come from large $|\phi |$, so that, after splitting the integration path into positive and negative regions, one can find two series expansions
for the fraction in (\ref{shift1}). The leading term for small $x$ is given by the positive $\psi$ part, that is
\begin{equation}
\begin{split}
-2&e^{-i\frac{\pi}{4}\left(1+\frac{1}{n}\right)}e^{\frac{\theta}{2n}}\int_{0}^{\infty} d\phi\hspace{.5cm} e^{\frac{\phi}{2}\left(1-\frac{1}{n}\right)-xE_{\phi}}+\ldots=\\
-2&e^{-i\frac{\pi}{4}\left(1+\frac{1}{n}\right)}e^{\frac{\theta}{2n}}\int_{-\infty}^{\infty} d\phi\hspace{.5cm} e^{-mx\cosh\phi}\cosh\left[\frac{\phi}{2}\left(1-\frac{1}{n}\right)\right]+\ldots=\\
-2&e^{-i\frac{\pi}{4}\left(1+\frac{1}{n}\right)}e^{\frac{\theta}{2n}}2^{\frac{1}{2}\left( 1-\frac{1}{n}\right)}\Gamma\left(\frac{n-1}{2n}\right) (mx)^{-\frac{1}{2}\left(1-\frac{1}{n}\right)}+\ldots.
\end{split}
\label{leadingterm1}
\end{equation}
The one particle form factor for $:\psi\T:$ can be read out from (\ref{0OTtheta2}) and (\ref{leadingterm1})
\begin{equation}
F_{1}^{:\psi\T:|1}(\theta)=-\frac{ie^{-i\frac{\pi}{4}\left(1+\frac{1}{n}\right)}}{\pi}2^{\frac{1}{2}\left( 1-\frac{1}{n}\right)}\Gamma\left(\frac{n-1}{2n}\right)\cos\left(\frac{\pi}{2n}\right)\vev{\T}\frac{m^{\frac{1}{2n}}}{\sqrt{4\pi}}
e^{\frac{\theta}{2n}}
\label{psiT2pff}
\end{equation}
The procedure used to get this result was introduced for descendants of twist fields in \cite{benjamin1}. In that case the authors were dealing with the twist field associated to the
global $U(1)$ symmetry of the Dirac Lagrangian. In this and next sections extensive use of this procedure is made, demonstrating its consistency for a different kind of twist fields.

\subsection{Two particle form factor of $:\E\T:$}
One of the main focuses of this paper is the computation of form factors of composite operators defined as in (\ref{definition}) with $\mathcal{O}=\E$.
The matrix element $\bra{0}\Theta(x)\T(0)\ket{\theta_{1}\theta_{2}}$ can be written by means of (\ref{csum}) as
\begin{equation}
\bra{0}\E(x)\T(0)\ket{\theta_{1},\theta_{2}}=\frac{1}{4\pi^{2}}\sum_{j=i}^{n}\int_{\phi_{1}>\phi_{2}} d\phi_{1}d\phi_{2}\hspace{.5cm}\bra{0}\E(x)\ket{\phi_{1},\phi_{2}}_{j,j}\,
   {\,}_{j,j}\!\bra{\phi_{2},\phi_{1}}\T(0)\ket{\theta_{1},\theta_{2}},
\label{et1}
\end{equation}
where $j$ labels the copy number, and is repeated because $\E$ connects only particles on the same copy.\\
The matrix element$ {\,}_{j,j}\!\bra{\phi_{2},\phi_{1}}\T(0)\ket{\theta_{1},\theta_{2}}$ is connected to the four-particle form factor of the twist field by the crossing relation (\ref{crossing}), which used repeatedly gives
\begin{equation}
\begin{split}
{\,}_{j,j}\!\bra{\phi_{2}\phi_{1}}\T(0)\ket{\theta_{1}\theta_{2}}&=-\bra{0}\T(0)\ket{\theta_{1},\theta_{2},\phi_{1}+i\pi-i\E^{+},\phi_{2}+i\pi-i\E^{+}}_{1,1,j,j}+\\
+\delta(\theta_{1}-\phi_{1})\delta_{1,j}\bra{0}\T(0)&\ket{\theta_{2},\phi_{2}+i\pi-i\E^{+}}_{1,j}+\delta(\theta_{2}-\phi_{2})\delta_{1,j}\bra{0}\T(0)\ket{\theta_{1},\phi_{1}-i\pi-i\E^{+}}_{1,j}-\\
\delta(\theta_{1}-\phi_{2})\delta_{1,j}\bra{0}\T(0)&\ket{\theta_{2},\phi_{1}+i\pi-i\E^{+}}_{1,j}-\delta(\theta_{2}-\phi_{1})\delta_{1,j}\bra{0}\T(0)\ket{\theta_{1},\phi_{2}-i\pi-i\E^{+}}_{1,j}+\\
+\delta(\theta_{1}-\phi_{1})\delta(\theta_{2}-\phi_{2})&\delta_{1,j}-\delta(\theta_{1}-\phi_{2})\delta(\theta_{2}-\phi_{1})\delta_{1,j}.
\end{split}
\label{crossing2}
\end{equation}
Even if (\ref{crossing2}) is quite cumbersome one can notice that many terms can be extracted from the others with the exchange $\theta_{1}\longleftrightarrow\theta_{2}$, which leaves
the integration untouched. The leading contribution for small $x$ is given by the first term (the one without deltas). This four-particle form factor considers two particles on the
first copy, and two on copy $j$, but can be reduced to one where particles are considered on the same copy (say $1$) with multiple applications of equations (\ref{1}) and (\ref{2})
\begin{equation}
\begin{split}
\bra{0}\T(0)&\ket{\theta_{1},\theta_{2},\phi_{1}+i\pi-i\E^{+},\phi_{2}+i\pi-i\E^{+}}_{1,1,j,j}=\\&F_{4}^{\T|11jj}(\theta_{1},\theta_{2},\phi_{1}+i\pi-i\E^{+},\phi_{2}+i\pi-i\E^{+})=\\
&F_{4}^{\T|1111}(\theta_{1},\theta_{2},\phi_{1}+(2j-1)i\pi,\phi_{2}+(2j-1)i\pi)_{+},
\end{split}
\label{leading2}
\end{equation}
and this allows us to use the Pfaffian structure (\ref{f}) to re-express it. In the last step of (\ref{leading2}) the notation $(\ldots)_{+}$ is introduced to indicate that any pole on the
real axe of what is in the brackets is avoided with the $i\E^{+}$ prescription. With the help of (\ref{f}), (\ref{k}) and (\ref{e2ff}), one can rewrite the leading term of (\ref{et1}) as
\begin{equation}
\begin{split}
\frac{ i m}{\vev{\T}} \frac{1}{4\pi^{2}}&\sum_{j=1}^{n}\int_{\phi_{1}>\phi_{2}} d\phi_{1}d\phi_{2}\hspace{.5cm} \sinh\left(\frac{\phi_{1}-\phi_{2}}{2}\right) e^{-ix(p_{\phi_{1}}+p_{\phi_{2}})}\\
&[ F_{2}^{\T|11}(\theta_{1},\theta_{2}) F_{2}^{\T|11}(\phi_{1},\phi_{2})-F_{2}^{\T|11}(\theta_{1},\phi_{1}+(2j-1)i\pi) F_{2}^{\T|11}(\theta_{2},\phi_{2}+(2j-1)i\pi)+\\
&F_{2}^{\T|11}(\theta_{2},\phi_{1}+(2j-1)i\pi) F_{2}^{\T|11}(\theta_{1},\phi_{2}+(2j-1)i\pi)  ]_{+},
\label{leading3}
\end{split}
\end{equation}
and proceed by evaluating the leading contribution for the three terms in (\ref{leading3}). The first one is
\begin{equation}
-\frac{m\cos\frac{\pi}{2n}}{8\pi^2}F_{2}^{\T|11}(\theta_{1},\theta_{2})\int d\phi_{1}d\phi_{2}\hspace{.5cm} \frac{\sinh\left( \frac{\phi_{1}-\phi_{2}}{2} \right)\sinh\left( \frac{\phi_{1}-\phi_{2}}{2n} \right)}{\sinh\left( \frac{\phi_{1}-\phi_{2}+i\pi}{2} \right)\sinh\left( \frac{\phi_{1}-\phi_{2}-i\pi}{2n} \right)}e^{-ix(p_{\phi_{1}}+p_{\phi_{2}})},
\label{firstterm}
\end{equation}
on which we can perform the change of variables $t=(\phi_{1}-\phi_{2})/2$ and $s=(\phi_{1}+\phi_{2})/2$, and carry out the $s$ integration to get
\begin{equation}
-\frac{m\cos\frac{\pi}{2n}}{2\pi^2}F_{2}^{\T|11}(\theta_{1},\theta_{2})\int dt\hspace{.5cm} \frac{\sinh t \sinh \frac{t}{n}}{\sinh\left( \frac{t}{n}+\frac{i\pi}{2n} \right)\sinh\left( \frac{t}{n}-\frac{i\pi}{2n} \right)}K_{0}(2mx\cosh t),
\label{firstterm2}
\end{equation}
We notice again that the leading contribution is given for large $t$ so that we can use the parity of the integrand to reduce the region of integration to $(0,\infty)$, and expand
the fraction in (\ref{firstterm2}) in a convergent way on this domain. The resulting integral is
\begin{equation}
-\frac{m\cos\frac{\pi}{2n}}{\pi^2}F_{2}^{\T|11}(\theta_{1},\theta_{2})\int_{0}^{\infty} dt\hspace{.5cm} e^{t\left(1-\frac{1}{n}\right)} K_{0}(mxe^{t})+\ldots,
\label{firstterm3}
\end{equation}
and with the change of variable $u=mxe^{t}$ we can extract the leading order for small $x$, that is
\begin{equation}
-\frac{m\cos\frac{\pi}{2n}}{\pi^2}F_{2}^{\T|11}(\theta_{1},\theta_{2})\left(\int_{0}^{\infty} dt\hspace{.5cm} u^{-\frac{1}{n}} K_{0}(u)\right)(mx)^{\frac{1}{n}-1}+\ldots.
\label{firstterm4}
\end{equation}
Solving the integral we finally obtain
\begin{equation}
-\frac{m\cos\frac{\pi}{2n}}{2^{1+\frac{1}{n}}\pi^2}\Gamma\left(\frac{n-1}{2n} \right)^{2}F_{2}^{\T|11}(\theta_{1},\theta_{2})(mx)^{\frac{1}{n}-1 }+\ldots.
\label{firstterm5}
\end{equation}
The second and third terms in (\ref{leading3}) are more involved then the first, due to their explicit dependence on the parameter $j$. One can start by noticing that the value of the third
term can be extracted from the value of the second one by changing sign, and performing the exchange $\theta_{1}\longleftrightarrow \theta_{2}$. Hence we focus on the first term. As, in order to
have a convergent integral, the integration axe has to be risen by $i\pi/2$, we need to study the pole structure of this term. The only kinematic poles which lie on the real axes arise
 from the cases $j=1,n$ and $\phi_{1}=\theta_{1}$ and $\phi_{2}=\theta_{2}$; but they are avoided with the $i\E^{+}$ prescription. In general $F_{2}^{\T|11}(\theta,\phi+(2j-1)i\pi)$ has
kinematic poles for $\phi=\theta+2(n-j+1)i\pi$ and $\phi=\theta+2(n-j)i\pi$. Considering that $j$ runs from $1$ to $n$ one can see that all poles group in even multiples of $i\pi$, so
that the first group above the real axe is in $\theta+2i\pi$, and correspond to $j=n,n-1$. Hence the needed shift can be safely performed.\\
The integral we want to evaluate is then
\begin{equation}
\begin{split}
\frac{im\vev{\T}\cos^{2}\frac{\pi}{2n}}{8\pi^2 n^{2}}\sum_{j=1}^{n}\int d\phi_{1}d\phi_{2}\hspace{.5cm}e^{-ix(p_{\phi_{1}}+p_{\phi_{2}})}\sinh\left( \frac{\phi_{1}-\phi_{2}}{2} \right)\\
\frac{\sinh\left( \frac{\theta_{1}-\phi_{1}-(2j-1)i\pi}{2n} \right)\sinh\left( \frac{\theta_{2}-\phi_{2}-(2j-1)i\pi}{2n} \right)}{\sinh\left( \frac{\theta_{1}-\phi_{1}-2(j-1)i\pi}{2n} \right)\sinh\left( \frac{\theta_{1}-\phi_{1}-2ji\pi}{2n} \right)\sinh\left( \frac{\theta_{2}-\phi_{2}-2(j-1)i\pi}{2n} \right)\sinh\left( \frac{\theta_{2}-\phi_{2}-2ji\pi}{2n} \right)}.
\end{split}
\label{secondterm1}
\end{equation}
Although the fraction in (\ref{secondterm1}) is rather complicated and mixes integration variables with parameters it can be dramatically simplified by means of the following identity
\begin{equation}
\begin{split}
&\sinh\left(\alpha_{1}-\beta_{1}\pm \gamma\right)\sinh\left(\alpha_{2}-\beta_{2}\pm \gamma\right)=\\
&\frac{1}{2}\left[ \cosh\left( \alpha_{1}+\alpha_{2}-(\beta_{1}+\beta_{2})\pm 2\gamma\right)-\cosh\left( \alpha_{1}-\alpha_{2}-(\beta_{1}-\beta_{2}\right)  \right],
\end{split}
\label{trigtrig}
\end{equation}
leading to
\begin{equation}
\begin{split}
\frac{im\vev{\T}\cos^{2}\frac{\pi}{2n}}{4\pi^2 n^{2}}\sum_{j=1}^{n}\int d\phi_{1}d\phi_{2}\hspace{.5cm}e^{-ix(p_{\phi_{1}}+p_{\phi_{2}})}\sinh\left( \frac{\phi_{1}-\phi_{2}}{2} \right)\\
\frac{\cosh \left( \frac{\theta_{1}+\theta_{2}-(\phi_{1}+\phi_{2})-2(2j-1)i\pi}{2n} \right)-\cosh\left( \frac{\theta_{1}-\theta_{2}-(\phi_{1}-\phi_{2})}{2n} \right)}{ \cosh \left( \frac{\theta_{1}+\theta_{2}-(\phi_{1}+\phi_{2})-4(j-1)i\pi}{2n} \right)-\cosh\left( \frac{\theta_{1}-\theta_{2}-(\phi_{1}-\phi_{2})}{2n} \right)}\times\\
\frac{1}{\cosh \left( \frac{\theta_{1}+\theta_{2}-(\phi_{1}+\phi_{2})-4ji\pi}{2n} \right)-\cosh\left( \frac{\theta_{1}-\theta_{2}-(\phi_{1}-\phi_{2})}{2n} \right)}.
\end{split}
\label{secondterm2}
\end{equation}
Now performing the same change of variable as in (\ref{firstterm2}) gives
\begin{equation}
\begin{split}
\frac{im\vev{\T}\cos^{2}\frac{\pi}{2n}}{2\pi^2 n^{2}}\sum_{j=1}^{n}\int dtds\hspace{.5cm}e^{-2imx\sinh s \cosh t}\sinh t\\
\frac{\cosh \left( \frac{\theta_{1}+\theta_{2}}{2n}-\frac{s+(2j-1)i\pi}{n} \right)-\cosh\left( \frac{\theta_{1}-\theta_{2}}{2n}-\frac{t}{n} \right)}{ \cosh \left( \frac{\theta_{1}+\theta_{2}}{2n}-\frac{s+2(j-1)i\pi}{n} \right)-\cosh\left( \frac{\theta_{1}-\theta_{2}}{2n}-\frac{t}{n} \right)}\times\\
\frac{1}{\cosh \left( \frac{\theta_{1}+\theta_{2}}{2n}-\frac{s-2ji\pi}{n} \right)-\cosh\left( \frac{\theta_{1}-\theta_{2}}{2n}-\frac{t}{n} \right)},
\end{split}
\label{secondterm3}
\end{equation}
and to make it convergent the shift $s\longrightarrow s-i\pi/2$ has been performed.\\
As $x$ approaches $0$ the main contribution is given by large $t$, and $s$ peaked around $0$. It is natural then to expand the fraction in (\ref{secondterm3}) in powers of $t$. As
before there is no such expansion on the whole real axe, but splitting it into the positive and negative parts, allows us to consider two different series which converge respectively on
the two regions. We can start considering $t<0$, such that the expansion of (\ref{secondterm3}) yields
\begin{equation}
\begin{split}
e^{-\frac{\theta_{1}-\theta_{2}}{2n}}\int_{-\infty}^{0}dt\hspace{.5cm}e^{-t\left(1-\frac{1}{n} \right)} \int_{-\infty}^{\infty} ds \hspace{.5cm}e^{-2mxe^{-t}\cosh s}+\ldots=\\
2e^{-\frac{\theta_{1}-\theta_{2}}{2n}}\int_{-\infty}^{0}dt\hspace{.5cm}e^{-t\left(1-\frac{1}{n} \right)}K_{0}(mxe^{-t})+\ldots.
\end{split}
\label{positivet}
\end{equation}
Following now the same procedure which was used to obtain (\ref{firstterm5}) out of (\ref{firstterm3}), we arrive at the result
\begin{equation}
\frac{im\vev{\T}\cos^{2}\frac{\pi}{2n}}{2^{\frac{1}{n}+1}\pi^2 n}\Gamma\left( \frac{n-1}{2n}\right)^{2}e^{-\frac{\theta_{1}-\theta_{2}}{2n}}(mx)^{\frac{1}{n}-1}+\ldots.
\label{firstresult}
\end{equation}
The positive part of the integral can be performed with the same logic and gives
\begin{equation}
-\frac{im\vev{\T}\cos^{2}\frac{\pi}{2n}}{2^{\frac{1}{n}+1}\pi^2 n} \Gamma\left( \frac{n-1}{2n}\right)^{2}e^{\frac{\theta_{1}-\theta_{2}}{2n}}(mx)^{\frac{1}{n}-1}+\ldots,
\label{secondresult}
\end{equation}
so that the final result of the second part is
\begin{equation}
-\frac{im\vev{\T}\cos^{2}\frac{\pi}{2n}}{2^{\frac{1}{n}}\pi^2 n}\Gamma\left( \frac{n-1}{2n}\right)^{2} \sinh \frac{\theta_{1}-\theta_{2}}{2n}(mx)^{\frac{1}{n}-1}+\ldots.
\label{secondresult2}
\end{equation}
As mentioned before, the result of of third part of the integral in (\ref{leading3}) can be obtained from (\ref{secondresult2}) by switching $\theta_{1}\longleftrightarrow\theta_{2}$ with a
minus sign in front, which doubles the result.
Putting (\ref{firstterm5}), (\ref{firstresult}) and (\ref{secondresult2}) together, we finally obtain the two particle form factor for the field $:\E\T:$, that is
\begin{equation}
F_{2}^{:\E\T:|11}(\theta_{1},\theta_{2})=-\frac{\cos\frac{\pi}{2n}}{2^{1+\frac{1}{n}}\pi^2}\Gamma\left(\frac{n-1}{2n} \right)^{2}m^{\frac{1}{n}}\left[F_{2}^{\T|11}(\theta_{1},\theta_{2})+
\frac{4i\cos\frac{\pi}{2n}}{n} \vev{\T}  \sinh \frac{\theta_{1}-\theta_{2}}{2n}\right].
\label{2pffet}
\end{equation}
This result was the aim of this section.
To check its validity one can employ (\ref{3}). Indeed, since $:\E\T:$ is still a twist field, the same type of residue equations as for $\T$ must be satisfied. Then one
can easily check using (\ref{vevs}) and (\ref{2pffet}) that
\begin{equation}
 \lim_{\bar{\theta}\rightarrow \theta}(\bar{\theta}- \theta)F_2^{:\E\T:|11}(\bar{\theta}+i\pi,\theta)=i\vev{:\E\T:},
\end{equation}
which confirms the compatibility of the two results of these sections.\\
Notice that this result satisfies all form factors equations for the twist field, and has a structure of the type
\begin{equation}
 \label{kernel}
F_2^{:\E\T:|11}(\theta_1,\theta_2)=\alpha \left[ Q_2^{\T}(\theta_1,\theta_2)+\beta\kappa(\theta_1,\theta_2) \right] F_{\text{min}}(\theta_1,\theta_2),
\end{equation}
where $\alpha$ and $\beta$ are two dimensional constants, $F_{\text{min}}$ is the minimal form factor of the theory\footnote{for the Ising model $F_{\text{min}}(\theta_1,\theta_2)=-i
\sinh \frac{\theta_1-\theta_2}{2n}$}, and $\kappa$ is a kernel solution of the form factor equations (see e.g. \cite{us1} for a discussion). Even if we could have understood that (\ref{2pffet}) should have had the form
(\ref{kernel}) to fulfill all twist properties it has to, we would have never been able to fix $\alpha$ and $\beta$ without the methods used in this section, and Section \ref{OPE}.

%

\section{Higher particle form factors of composite operators}
\label{higher}
In this section we deal with the computation of higher particle form factors for $:\psi\T:$ and $:\E\T:$. They can be extracted with the same methods for both operators, employing
higher particle form factors of the twist field (\ref{f}). Indeed, focusing on $:\psi\T:$, when looking for the leading term for the $2k-1$ particle form factor, one has to deal with
\begin{equation}
 \bra{0}\psi(x)\T(0)\ket{\theta_1,\theta_2,\dots,\theta_{2k-1}}\sim \frac{n}{2\pi}\int d\phi\bra{0}\psi(x)\ket{\phi}
\bra{0}\T(0)\ket{\theta_1,\theta_2,\dots,\theta_{2k-1},\phi+i\pi-i\E^{+}}.
\end{equation}
One is then able to isolate term by term the higher particle part, and reduce it to the same evaluation carried out in (\ref{0OTtheta})-(\ref{psiT2pff}), getting finally
\begin{equation}
 F_{2k-1}^{:\psi\T:|11...1}=\vev{\T}{\text{Pf}}(K_{:\psi\T:}),
\end{equation}
where $K_{:\psi\T:}$ is the $2k\times2k$ matrix defined as
\begin{equation}
K_{:\psi\T:}=\begin{pmatrix}
  0 & F_1^{:\psi\T:|1}(\theta_1) & \cdots & F_{1}^{:\psi\T:|1}(\theta_{2k-1}) \\
  -F_1^{:\psi\T:|1}(\theta_1) & 0 & \cdots & F_2^{\T|11}(\theta_{1},\theta_{2k-1})/\vev{\T} \\
  \vdots  & \vdots  & \ddots & \vdots  \\
  -F_{1}^{:\psi\T:|1}(\theta_{2k-1}) & -F_2^{\T|11}(\theta_{1},\theta_{2k-1})/\vev{\T} & \cdots & 0
 \end{pmatrix}.
\end{equation}
Higher particles form factors for $:\E\T:$ can be evaluated with the same logic, although they show a more complicated pattern, and can not be reduced to a Pfaffian form. This is due
to the presence of a kernel part in the two particle form factor. The $2k+2$ particle form factor is
\begin{equation}
 F_{2k+2}^{:\E\T:|11...1}(\theta_1,\theta_2,\dots,\theta_{2k+2})=\sum_{i<j}\frac{(-1)^{\sigma(i,j)}}{\vev{\T}^{2k}}F^{\T|11...1}_{2k}(\theta_1,\dots,\theta_{2k+2})_{ij}F_2^{:\E\T:|11}(\theta_i,\theta_j),
\end{equation}
where $\sigma(i,j)$ is the permutation that brings $\theta_i$ and $\theta_j$ to the right of all other rapidities, while with $F(\dots)_{ij}$ we mean a form factor of all rapidities
but those two.

\section{Conclusions}
\label{conclusions}
In this paper we have investigated the short-distance behaviour of the correlation function $\vev{\E(r)\T(0)}$ for the two dimensional Ising model in the vicinity of the critical point beyond the leading contribution. This led us to the identification of the vacuum expectation values of a new class of twist fields, including
\begin{equation}
:\E\T:(x)\sim \lim_{\delta\rightarrow0}\E(x+\delta)\T(x),
\end{equation} 
and its derivatives.
Furthermore we managed to compute massive corrections to the structure constants up to $(mr)^{6}$ for all these fields. The very fact that we are able to compute VEVs of such a large set of local operators is remarkable, as the computation of VEVs for general theories and fields is known to be a very hard task and there is no general procedure to tackle it.
In addition to the contributions that would be naturally expected from the CFT theory we have found logarithmic corrections, a phenomenon that has already been observed for other correlation functions in the off-critical Ising model (see e.g. \cite{spinspin}).
These logarithmic terms are due to the arbitrariness in the definition of $\vev{\E}$.
In addition we have computed all higher-particle form factors of $:\E\T:$ and $:\psi\T:$. By exploiting (\ref{definition}) we have been able to fully determine the normalization of all form factors and to provide new solutions to the form factor equations for twist fields. A byproduct of our investigation is the fact that all the expectation values $\vev{:\de^{2\alpha}\E\T:}$ are negative. Although we did not put much emphasis on this feature in the present paper, this provides further evidence for the negativity of the connected correlator $\vev{\E(r)\T(0)}$, claimed in \cite{us2}.\\

There are a number of open problems related to the present work which we would like to address in future. First, the short distance expansion of $\E(r)\T(0)$ that we considered here is only well defined for even $n$. It would be interesting to have a deeper understanding of what happens when we consider $n$ odd, and why these two cases are distinct. 
Second, this paper introduces an OPE which involves the twist field in a replica theory. This type of OPEs have never been studied, neither in the massive nor in the critical theories. A better understanding of the operator content and OPEs in  replica theories is desirable. Finally, it would be interesting to understand if these composite twist fields are in any way related to the entanglement entropy of particular states in the Ising model.\\    

\textbf{Acknowledgements}
The author wishes to thank Olalla Castro-Alvaredo for her guidance and for proofreading this manuscript (without her help the completion of this project would not have been possible), and Benjamin Doyon for precious explanations and discussions.\\

\appendix
\section{Coefficients $\Omega_\alpha(n)$}
\label{A}
This section is devoted to the computation of coefficients in (\ref{exp1}). As explained in Section \ref{OPE} the aim is to expand the fraction on the LHS of that equation for large
$t/mr$. First of all, let us introduce the more convenient variable $u=(t/mr)^{-2/n}$. The denominator can be then treated, as long as $t> mr$ as the generating function of the
Chebishev polynomials of the second kind, that is
\begin{equation}
 \label{cheb1}
\frac{1}{u^2-2xu+1}=\sum_{\alpha=0}^{\infty}U_\alpha(x)u^\alpha,
\end{equation}
for $-1<x<1$, and $|u|<1$. The first condition is satisfied for every $n\geq 2$ as $x=\cos(\pi/n)$, while the second is satisfied in the whole integration domain of (\ref{int2}) except for the lower limit $t=mr$. This divergence is
``cured'' by integrating over the domain $[mr+\epsilon,\infty)$, where $\epsilon$ is a small parameter. Once this expansion is plugged into (\ref{int2}) the sum and the integration can be safely exchanged. After the
integration is performed one has then to be sure that the result does not depend on $\epsilon$, and finally set it to zero. We have performed these steps showing that indeed (\ref{cheb1})
in this case can be taken as valid also at the point $t=mr$. The details are technical and cumbersome, and they will not be reported here. From now on, and throughout the calculation
in Section \ref{OPE} we will take (\ref{cheb1}) as series representation an the whole integration path.\\
The polynomials $U_\alpha(x)$ in our case are formally defined as follows
\begin{equation}
 \label{cheb2}
U_\alpha\left(\cos \frac{\pi}{n}\right)=\frac{\sin\frac{(1+\alpha)\pi}{n}}{\sin \frac{\pi}{n}},
\end{equation}
The LHS in (\ref{exp1}) can then be expanded as shown in the RHS with
\begin{equation}
 \label{coeffcheb}
\Omega_\alpha(n)= \left\{
  \begin{array}{l l}
    \frac{\cos\frac{(1+2\alpha)\pi}{2n}}{\cos\frac{\pi}{2n}}  \quad &\text{if $\alpha<n$}\\
   \frac{\cos\frac{(1+2\alpha)\pi}{2n}}{\cos\frac{\pi}{2n}} + \frac{\sin\frac{(1+2\alpha)\pi}{2n}}{\sin\frac{\pi}{2n}}\quad &\text{if $ \alpha \geq n$}\\
\end{array} \right. .
\end{equation}

\section{Definite integrals of Bessel functions and powers}
\label{B}
In this appendix we present a solution to integrals of the kind
\begin{equation}
\label{genint}
\int_{mr}^{\infty}dt\hspace{.3cm}t^{-\mu}K_{\nu}(t),
\end{equation}
where both $\mu$ and $mr$ are positive real numbers. In Section \ref{OPE}, in particular, an expansion for small values of $mr$ was needed, so that this will be the aim of this appendix.
First of all let us introduce the function
\begin{equation}
\label{maijer}
G_{p,q}^{\,m,n} \!\left(    \, t \left| \begin{matrix} a_1, \dots, a_p \\ b_1, \dots, b_q \end{matrix}\;\right.  \right) = \frac{1}{2 \pi i} \int_L \frac{\prod_{j=1}^m \Gamma(b_j - s) \prod_{j=1}^n \Gamma(1 - a_j +s)} {\prod_{j=m+1}^q \Gamma(1 - b_j + s) \prod_{j=n+1}^p \Gamma(a_j - s)} t^s \,ds.
\end{equation}
This is a representation of the Meijer $G$-function, in the formalism adopted by \cite{table}, and the details and properties about this function will not be reported here.
A useful identity is
\begin{equation}
\label{besselmaijer}
K_\nu (t) = \frac{1}{2} \; G_{0,2}^{\,2,0} \!\left(  \, \frac{t^2}{4}\left| \frac{\nu}{2}, \frac{-\nu}{2} \; \right. \right),
\end{equation}
which holds for $|\arg(t)|\leq \pi/2$, and the empty sets of gamma functions' poles are omitted. In the light of (\ref{besselmaijer}) the integral in (\ref{genint}) can be rewritten as follows
\begin{equation}
\label{genint2}
\frac{(mr)^{1-\mu}}{4}\int_{1}^{\infty}dt \hspace{.3cm}t^{-\frac{\mu+1}{2}}G_{0,2}^{\,2,0} \!\left(  \, \frac{m^2}{4}t\left| \frac{\nu}{2}, -\frac{\nu}{2} \; \right. \right).
\end{equation}
This is a special case of a known integral of the $G$-function, which in the most general form is
\begin{equation}
\label{genintG}
\begin{split}
\int_{1}^{\infty}dt\hspace{.3cm}t^{-\rho}(t-1)^{\sigma-1}G_{p,q}^{\,m,n} \!\left(    \, \alpha t \left| \begin{matrix} a_1, \dots, a_p \\ b_1, \dots, b_q \end{matrix}\;\right.  \right)
=\Gamma(\sigma)G_{p+1,q+1}^{\,m+1,n} \!\left(    \, \alpha \left| \begin{matrix} a_1, \dots, a_p ,\rho\\ \rho-\sigma,b_1, \dots, b_q \end{matrix}\;\right.  \right),
\end{split}
\end{equation}
which holds for real $|\arg(t)|\leq (m+n-p/2-q/2)\pi$, $p+q\leq2(m+n)$, $\Re(\sigma)>0$ and $\Re(\rho-\sigma-a_{j})>-1$ $\forall j\in[1,n]$. These conditions are all clearly satisfied by (\ref{genint2}), so that the result is
\begin{equation}
\label{resultGfunction}
\int_{mr}^{\infty}dt\hspace{.3cm}t^{-\mu}K_{\nu}(t)=\frac{(mr)^{1-\mu}}{4}G_{1,3}^{\,3,0}\!\left(    \, \frac{m^2}{4} \left| \begin{matrix}  \frac{\mu+1}{2}  \\ \frac{\mu-1}{2}\; , \frac{\nu}{2}\; , -\frac{\nu}{2} \end{matrix}\;\right.  \right)
\end{equation}
Now this result has to be restricted to the cases (\ref{besselexp}) to be useful for the OPE that was considered in Section \ref{OPE}. The calculations are tedious and the results cumbersome, hence only the first few terms of the expansion of the first two contributions are given
\begin{equation}
\label{finalexpansion1}
\begin{split}
&\int_{mr}^{\infty}dt\hspace{.3cm}t^{-\frac{2\alpha+1}{n}}K_{0}(t)=2^{-1-\frac{1+2\alpha}{n}}\Gamma\left( \frac{n-1-2\alpha}{2n}\right)^{2}+\\
&+\frac{n[(n-1-2\alpha)(\gamma-\log2)-n]}{(n-1-2\alpha)^{2}}(mr)^{1-\frac{1+2\alpha}{n}}+\frac{n}{(n-1-2\alpha)}(mr)^{1-\frac{1+2\alpha}{n}}\log (mr)+\\
&+\frac{n[(3n-1-2\alpha)(\gamma-1-\log2)-n]}{4(3n-1-2\alpha)^{2}}(mr)^{3-\frac{1+2\alpha}{n}}+\frac{n}{4(3n-1-2\alpha)}(mr)^{3-\frac{1+2\alpha}{n}}\log (mr)+\\
&+O(mr)^{5-\frac{1+2\alpha}{n}}
\end{split}
\end{equation}

\begin{equation}
\label{finalexpansion2}
\begin{split}
&\int_{mr}^{\infty}dt\hspace{.3cm}t^{-1-\frac{2\alpha+1}{n}}K_{1}(t)=\frac{(mr)^{-1-\frac{1+2 \alpha}{n}} n}{1+2 \alpha+n}+2^{-2-\frac{1+2\alpha}{n}}
\Gamma\left( \frac{n-1-2\alpha}{2n}\right)\Gamma\left( -\frac{1+2\alpha+n}{2n}\right)+\\
&\frac{n[(n-1-2\alpha)(1+2\log 2-2\gamma)-2n]}{4(n-1-2\alpha)^{2}}(mr)^{1-\frac{1+2\alpha}{n}}-\frac{n}{2(n-1-2\alpha)}(mr)^{1-\frac{1+2\alpha}{n}}\log (mr)+\\
&+\frac{n[(3n-1-2\alpha)(5+4\log2-4\gamma)+4n]}{64(3n-1-2\alpha)^{2}}(mr)^{3-\frac{1+2\alpha}{n}}+\frac{n}{16(3n-1-2\alpha)}(mr)^{3-\frac{1+2\alpha}{n}}\log (mr)
+\\
&+O(mr)^{5-\frac{1+2\alpha}{n}}
\end{split}
\end{equation}
where $\gamma=0.577216$ is the Euler-Mascheroni constant.

\end{document}